\documentclass[12pt]{iopart}

\expandafter\let\csname equation*\endcsname\relax
\expandafter\let\csname endequation*\endcsname\relax
\usepackage{iopams}  
\usepackage{graphicx}
\usepackage{physics}
\usepackage{romannum}
\usepackage{hyperref}
\usepackage{siunitx}
\usepackage{subfigure}

\begin{document}

\title[Absorptive LTM: combining visible diamond Raman lasers and NV centres]{Absorptive laser threshold magnetometry: combining visible diamond Raman lasers and nitrogen-vacancy centres}
\author{Sarath Raman Nair$^{1, 2, *}$, Lachlan J. Rogers$^{1, 2}$, David J. Spence$^{1}$, Richard P. Mildren$^{1}$, Fedor Jelezko$^{3}$, Andrew D. Greentree$^{4}$, Thomas Volz$^{1, 2}$, and Jan Jeske$^{5}$}
\address{
 $^{1}$Department of Physics and Astronomy, Macquarie University, NSW 2109, Australia}
\address{
 $^{2}$ARC Centre of Excellence for Engineered Quantum Systems, Macquarie University, NSW 2109, Australia}
\address{
$^{3}$Institute for Quantum Optics, Ulm University, Albert-Einstein-Allee 11, D-89081, Ulm, Germany}
\address{
$^{4}$ARC Centre of Excellence for Nanoscale BioPhotonics, School of Science, RMIT University, Melbourne, VIC, 3001, Australia}
\address{
 $^{5}$Fraunhofer Institut f{\"u}r Angewandte Festk\"orperphysik IAF, Tullastrasse 72, 79108 Freiburg, Germany}
\address{}
\address{$^{*}$Corresponding author}
\ead{sarath.raman-nair@mq.edu.au}

\begin{abstract}
We propose a high-sensitivity magnetometry scheme based on a diamond Raman laser with visible pump absorption by an ensemble of coherently microwave driven negatively charged nitrogen-vacancy centres (NV) in the same diamond crystal.
The NV centres' absorption and emission are spin-dependent. We show how the varying absorption of the NV centres changes the Raman laser output. A shift in the diamond Raman laser threshold and output occurs with the external magnetic field and microwave driving. 
We develop a theoretical framework including steady-state solutions to describe the effects of coherently driven NV centres in a diamond Raman laser.
We discuss that such a laser working at the threshold can be employed for magnetic field sensing.
In contrast to previous studies on NV magnetometry with visible laser absorption, the laser threshold magnetometry method is expected to have low technical noise, due to low background light in the measurement signal. 
For magnetic-field sensing, we project a shot-noise limited DC sensitivity of a few pT/$\sqrt{\rm{Hz}}$ in a well-calibrated cavity with realistic parameters. 
This sensor employs the broad visible absorption of NV centres and unlike previous laser threshold magnetometry proposals it does not rely on active NV centre lasing or an infrared laser medium at the specific wavelength of the NV centre's infrared absorption line.
\end{abstract}

%
%
%
%
%
\section{\label{sec:level1}Introduction}

Negatively charged nitrogen-vacancy (NV) centres in diamond are being widely explored as quantum sensors for magnetic fields~\cite{doherty2013nitrogen, Rondin2014, degen2017quantum, barry2020sensitivity}, electric fields~\cite{Dolde2011, Dolde2014}, pressure~\cite{Doherty2014} and temperature~\cite{Neumann2013, degen2017quantum}.
They are also promising candidates for qubits~\cite{wrachtrup2006processing}, fluorescent biomarkers~\cite{McGuinness2011, Reineck2017, Miller2020}, nanoscale~\cite{maletinsky2012robust, Degen2008} and microscale~\cite{mizuno2018wide, McCloskey2020} magnetic field mapping, decoherence microscopy~\cite{Cole2009, Hall2009, Jeske2012}, nanoscale nuclear magnetic resonance~\cite{Mamin2013}, optical trapping~\cite{juan2017cooperatively}, laser-cooling~\cite{Kern2017} and gyroscopes~\cite{Ledbetter2012}.
For sensing, NV centres have the potential to achieve room-temperature operation enabling sensing without the need for insulation barriers between the sensor and sources such as biological tissue. 
Such complications are needed in other high-sensitivity technologies such as cryogenic superconducting quantum interference devices (SQUIDs) and vapour cell magnetometers.

Laser threshold magnetometry (LTM)~\cite{jeske2016laser, jeske2017StimEm, nair2020amplification, Fraczek2017, dumeige2019infrared} is a new concept of magnetic field sensing that harnesses the non-linearity of a laser cavity to strongly improve sensitivity to small magnetic field changes. 
Microwave (MW) driven negatively charged NV centres were proposed \cite{jeske2016laser} as an active laser gain medium, such that the laser threshold shifts with the external magnetic field. 
Small magnetic field changes then result in strong changes of the laser output power, leading (theoretically) to a huge improvement of the shot-noise limited DC sensitivities potentially down to the order of fT/$\sqrt{\mathrm{Hz}}$ for a crystal sensing volume of 1mm$^3$. Such sensitivity range would be several orders of magnitude better than even the state of the art AC magnetometry with NV centres \cite{wolf2015subpicotesla}. 
Although the stimulated emission of NV centres without an external cavity \cite{jeske2017StimEm} and also stimulated emission of NV centres and the amplification of an external laser inside an optical cavity \cite{nair2020amplification} have been shown in experiments, the challenge of achieving active NV lasing is still outstanding. 
A variation has been proposed \cite{dumeige2019infrared}, which does not require active NV lasing, but instead the idea is to use infrared (IR) singlet transition of the NV centre (1042~nm) as a variable absorber inside the cavity of another IR active gain medium, thus can shift the IR laser threshold with the external magnetic field.
This study predicted a theoretical shot-noise DC sensitivity around 700~fT/$\sqrt{\mathrm{Hz}}$. 
However, for this idea to work an intracavity IR laser medium needs to be matched with the relatively narrow absorption line upon which sensing is then based.
Moreover, the IR wavelength absorption by NV centre is weak compared to visible wavelength absorption.
Both are promising and recent concepts, which have not yet been experimentally realised.

Here we propose a new absorptive laser threshold magnetometry approach using the broad visible absorption spectrum of the NV centre as a magnetically controlled absorber~\cite{walsworth2017absorbtion, ahmadi2018nitrogen} of the pump wavelength inside the cavity of a diamond Raman laser.
Raman lasers rely on amplification of a cavity Stokes field through stimulated scattering of a pump by zone-centre phonons~\cite{loudon2001raman}.
Hence the threshold of a diamond Raman laser is influenced by the intensity of the pump field or the cavity quality factor at the Stokes frequency, and is dependent on absorption by impurities. 
As a result, traditional diamond Raman lasers require highly perfect diamond crystals, and so defect centres are seen as sources of scattering and loss.  
Investigating the effects of NV centres in diamond Raman lasing~\cite{mildren2013optical, mildren2014diamond, williams2018high} is an interesting new perspective for sensing using the concept of LTM.
Since the absorption of the pump by NV centres changes the threshold of the Raman laser, the Raman laser threshold in turn changes with external magnetic field.
We provide the first investigation of the effects of NV centres on diamond Raman lasers and our study shows that NV centres in diamond Raman crystal can instead be a resource for sensing based on the spin-dependent ground state population of these centres and its variation with magnetic field.
Compared to most NV magnetometry schemes our approach is based on the existing absorption-based magnetometry, which relies on the absorption of a visible external laser transmitting through the diamond with NV centres \cite{walsworth2017absorbtion, ahmadi2018nitrogen}.
This absorption-based scheme has opposite behaviour of bright and dark states: Upon transition to the ``darker'' m$_{\mathrm{s}}$=1 state the relatively long-lived singlet state is populated and thus the ground state population reduced.
This results in reduced NV absorption and thus stronger Raman laser output in our case. Thus the commonly ``darker'' m$_{\mathrm{s}}$=1 state, which shows less fluorescence, actually leads to higher signal output than the m$_{\mathrm{s}}$=0 spin state in the absorption-based scheme.
In contrast to the existing visible absorption-based magnetometry \cite{walsworth2017absorbtion, ahmadi2018nitrogen}, in which change in the laser transmission with magnetic field is of interest, our method working at a laser threshold has the advantage of significant reduction of background light, with which a low technical noise while measurements is expected.
The present concept allows the combination of NV centres with an established active laser system \cite{mildren2013diamond, mildren2014diamond, williams2018high}, which is also based on diamond crystals.
Comparing with two separate media in the IR LTM proposed in reference \cite{dumeige2019infrared}, the absorption and light generation in our present study can be realised in a single crystal.
The properties of the Raman gain mechanism offer a wide diversity of options for implementing the Raman laser: operation is compatible with almost any wavelength throughout the wide transmission band of diamond, overlapping with the broad visible absorption band, IR absorption, and both near IR as well as IR emission of the NV centre. 
The broad visible wavelength absorption, which is the focus in the present study, allows for multiple laser wavelengths in the visible range and improves robustness of the sensing concept against wavelength fluctuations relative to the narrow absorption of the infrared transition.
Laser operation is also compatible across a wide range of temporal regimes (continuous to ultrafast pulses)\cite{kitzler2012continuous, sabella20101240, spence2010mode, nikkinen2018sub, murtagh2015efficient}. 
Designs have been demonstrated in several physical formats including high-power free-space designs~\cite{ williams2018high}, efficient monolithic devices \cite{reilly2015monolithic} and miniature on-chip devices with low threshold ($<$ 80~mW \cite{latawiec2015chip}). 
Hence magnetic field sensors based on Raman lasing may take a variety of forms and operate across a wide range of scales.

This article is organized in three sections.
In Section \Romannum{1}, we model a continuous-wave (CW) diamond Raman laser in a Fabry-Perot cavity with intra-cavity absorption by MW driven negatively-charged NV centres in the same diamond crystal.
We show numerical results of the Raman laser output in the presence of an external magnetic field and its sensitivity in Section \Romannum{2}.
In Section \Romannum{3}, we conclude by discussing the implications of the results in actual experimental scenarios.

\section{\label{sec:level2}Model}
\begin{figure}
\centering
\includegraphics[scale = 0.8]{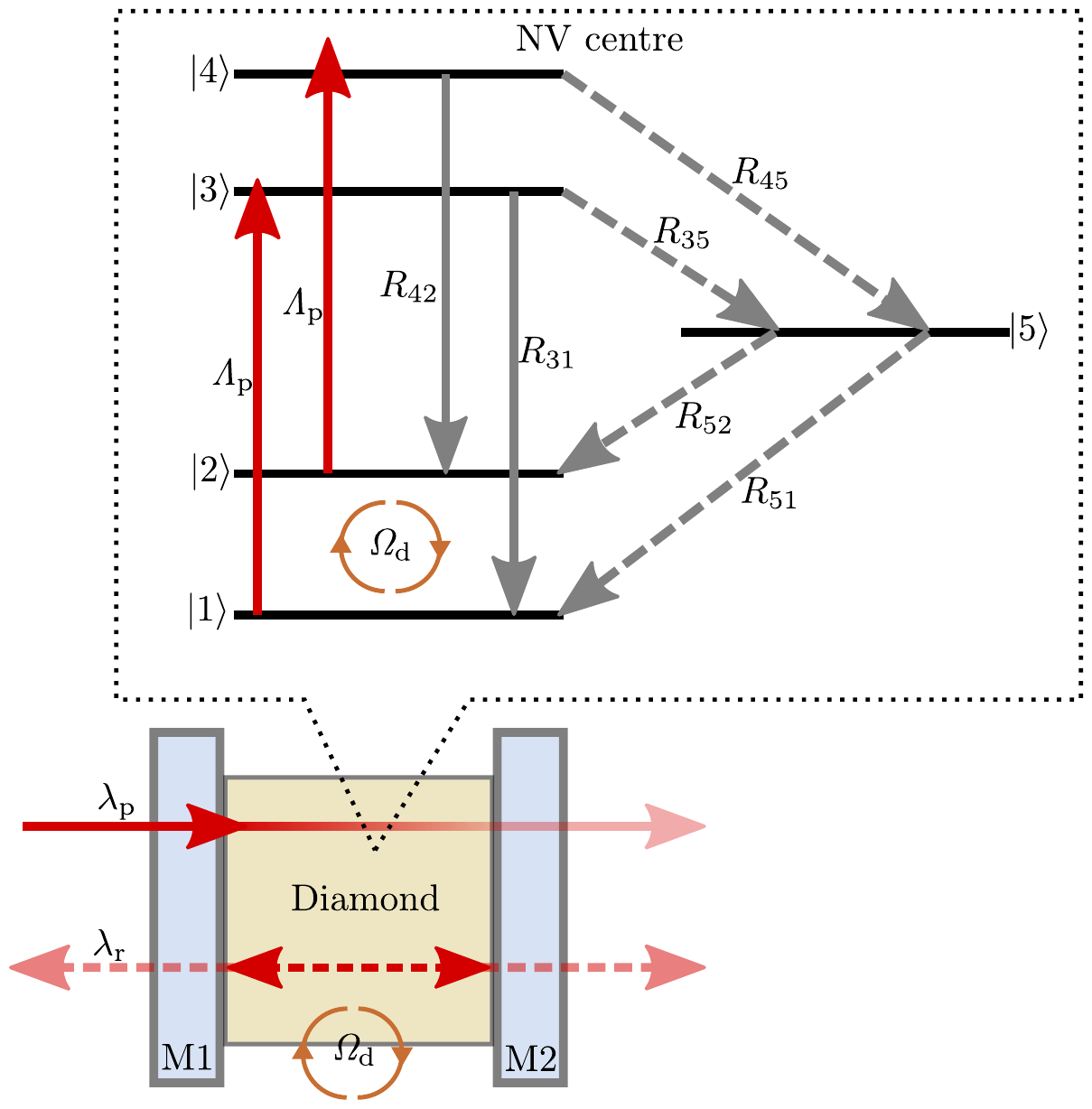}

\caption{Schematic of the Raman laser with a diamond crystal containing MW driven NV centres in the Fabry-Perot cavity.
The Fabry-Perot cavity is formed by two highly reflective mirrors denoted as M1 and M2.
Directions of the pump ($\lambda_{\mathrm{p}}$) and Raman ($\lambda_{\mathrm{r}}$) wavelengths are shown in solid and dotted red arrows respectively.
The pump is considered as a single pass pump, i.e. the cavity mirrors are not reflective for the pump wavelength.
The pump intensity depletes while propagating through the diamond and the light at the Raman wavelength is generated.
The Raman wavelength forms a cavity mode and a laser signal is generated. 
The laser output is transmitted through both mirrors.
A simplified level structure model of the MW driven NV centre present in the diamond is shown in the inset.
Solid red upward arrows represent the optical pumping of the NV centres by the $\lambda_{\mathrm{p}}$ wavelength.
The solid and dashed grey downward arrows represent radiative and non radiative (via the singlet state) decay paths.
Brown arrows denote the coherent MW driving of the NV centre with Rabi frquency $\varOmega_{\mathrm{d}}$.
The incoherent pumping rates ($\varLambda_{\mathrm{p}}$) and the decay rate for each transition are denoted next to the corresponding arrow.
}
\label{fig:1}
\end{figure}
We model CW diamond Raman laser system that works at ambient conditions, considering MW driven NV centres in the same diamond crystal as intra-cavity absorbers.
The visible absorption band of the NV centres, more precisely the diamond crystal containing NV centres spans from 500~nm to 700~nm.
The zero-phonon line (ZPL) of the NV centre is at 637~nm, which corresponds to around 471~THz.
Due to the characteristic diamond Raman shift, the Raman laser frequency is 40~THz lower than that of the pump light frequency, i.e.~the respective wavelength is higher.
NV centres can in principle absorb both pump and Raman wavelengths, if both these wavelengths fall into the absorption band of the centres.
Magnetic-field detection using the absorption by NV centres is possible only if there is a steady-state population in the NV centres' ground state available for spin manipulation.
The intense intra-cavity field of the Raman wavelength will quickly saturate the NVs to its excited state or singlet states, thus removing any potential for the variation of ground state population and absorption.
Furthermore  bistable behaviour above threshold can occur, since NV centres are saturable absorbers \cite{zhang2013laser,zhang2015large,seidel1971bistable}. 
Hence we consider the desired operating condition with pump wavelength in the absorption spectrum of the NV centres, i.e.~below the ZPL, and a Raman wavelength outside that absorption spectrum, i.e.\,above the ZPL.

For this study, we consider that the pump wavelength ($\lambda_{\mathrm{p}}$) is 620~nm (around 484~THz), which is at the upper end of the NV absorption spectrum, and thus the corresponding Raman wavelength ($\lambda_{\mathrm{r}}$) is around 676~nm (around 444~THz).
Though the tail of the experimental absorption band of diamond containing NV centres includes the Raman wavelength around 676~nm \cite{Kern2017}, we neglect the absorption of this wavelength by the NV centres for two reasons. 
First, the Raman laser photons are not expected to excite NV centres from their ground state to the excited state, since 
these photons are red-detuned by around 27~THz with respect to the ZPL of the NV centre. 
Second, we do not expect a significant steady-state population to occupy the phonon levels of the NV ground state, as they have extremely short lifetime on the order of picoseconds \cite{jeske2016laser}.
Hence, we consider that the NV centres absorb only the pump wavelength at 620~nm

We obtain a steady-state solution to the CW diamond Raman laser system incorporating MW driven NV centres at ambient conditions in two steps.
Firstly, we model ideal NV centres coherently driven by MW as the absorbers, in section (\ref{subsec:2.1}). 
Secondly, we model the diamond Raman laser including the absorbers to get a pump-laser output relation in section (\ref{subsec:2.2}).
The schematic of our model is shown in Figure \ref{fig:1}.

\subsection{\label{subsec:2.1}MW driven NV centres as the absorbers in the crystal}
We assume all the NV centres in the crystal are non-interacting  and identical absorbers.
NV centres are typically oriented along four different crystal directions. For simplicity, we work with only one orientation. This can be achieved in a crystal produced with preferential orientation \cite{edmonds2012production}.
At ambient conditions, each NV centre in the ensemble can be represented as a five-level system as shown in Figure \ref{fig:1} \cite{dumeige2019infrared, tetienne2012magnetic, barry2020sensitivity, doherty2013nitrogen}.
For simplicity, only one of the two spin  m$_{\mathrm{s}} = \pm$1 levels is considered, which can be achieved by lifting the degeneracy using Zeeman effect by applying a background magnetic field along the quantization axis and choosing a MW frequency only resonant with one transition. 
Levels $\ket{1}$ and $\ket{2}$ represent m$_{\mathrm{s}} =$ 0 and one of the m$_{\mathrm{s}} = \pm$1 levels of the ground state respectively, levels $\ket{3}$ and $\ket{4}$ represent the m$_{\mathrm{s}} =$ 0 and one of the m$_{\mathrm{s}} = \pm$1 levels of the excited state. We represent the singlet states as a single level $\ket{5}$.

The ground state levels $\ket{1}$ and $\ket{2}$ are resonantly driven by MW with a driving (Rabi) frequency $\varOmega_{\mathrm{d}}$.
We assume that the incoherent optical pumping of the NV centres with $\lambda_{\mathrm{p}}$ excites the NV centre from its ground state to excited state with pumping rate $\varLambda_{\mathrm{p}}$.
This incoherent optical excitation occurs via spin conserving transitions ($\Delta$m$_{\mathrm{s}}$ = 0) and hence excites the NV centre from states $\ket{1}$ and $\ket{2}$ to states $\ket{3}$ and $\ket{4}$ respectively.
These incoherent optical transitions in fact involve phonon levels of excited states $\ket{3}$ and $\ket{4}$, but since those phonon levels have extremely short lifetime, compared with that of $\ket{3}$ and $\ket{4}$, we neglect them for simplicity.
Since this incoherent laser pumping is spin conserving, the linewidth of the laser used is not critical for NV centre system.
We treat appreciable coherence only between the $\ket{1}$ and $\ket{2}$ and assume an intrinsic dephasing rate, $\varGamma_{\mathrm{g}}$ between these levels.
This intrinsic dephasing rate ($\varGamma_{\mathrm{g}}$) of the NV centre ground state is assumed to be 1~MHz unless specified otherwise \cite{doherty2013nitrogen,jeske2016laser}.
We also consider the effect of laser pumping $\varLambda_{\mathrm{p}}$ on this transition and neglect the spin relaxation since continuous laser and MW driving determine the spin dynamics on a faster time scale.

The NV centre population from $\ket{3}$ and $\ket{4}$ spontaneously decay to the ground state via two path ways \cite{doherty2013nitrogen, tetienne2012magnetic}.
The first one, involves spin-conserving optical transitions emitting near-IR photons, from $\ket{3}$ and $\ket{4}$ to $\ket{1}$ and $\ket{2}$, respectively.
The second one, via  non spin-conserving transitions from $\ket{3}$ and $\ket{4}$ to $\ket{1}$ and $\ket{2}$ through $\ket{5}$.
The transition rate corresponding to transition $\ket{i}$ to $\ket{j}$ is denoted as $R_{ij}$.
In this study, we consider NV transition rates as $R_{31} = R_{42}$= \SI{66.16}{{\micro\second}^{-1}}, $R_{45}$ = \SI{91.8}{{\micro\second}^{-1}}, $R_{35}$ = \SI{11.1}{{\micro\second}^{-1}}, $R_{51}$ = \SI{4.87}{{\micro\second}^{-1}}, $R_{52}$ = \SI{2.04}{{\micro\second}^{-1}} \cite{barry2020sensitivity, gupta2016efficient}. The non spin-conserving transition rates to and from the singlet state $\ket{5}$ are assumed constant although recently there have been indications that they are changing with magnetic field \cite{capelli2017magnetic}.

Since the absorption of the pump wavelength happens from both levels $\ket{1}$ and $\ket{2}$, their populations determine the intra-cavity absorption by NV centres per crystal length, which is relevant to the Raman laser model in the following section.
To obtain populations, the master equation for the density matrix $\rho$ of the NV centre from the 5-level system can be written as \cite{breuer2002theory, jeske2013derivation, jeske2016laser, vogt2013},
\begin{align}
\label{eq:1}
\dv{\rho_{12}}{t}&=(i \varDelta_{\mathrm{g}}-\varLambda_{\mathrm{p}}-\varGamma_{\mathrm{g}})\rho_{12}- i \frac{\varOmega_{\mathrm{d}}}{2}(\rho_{22}-\rho_{11}),\\
\label{eq:2}
\dv{\rho_{21}}{t}&=-(i \varDelta_{\mathrm{g}}+\varLambda_{\mathrm{p}}+\varGamma_{\mathrm{g}})\rho_{21}- i \frac{\varOmega_{\mathrm{d}}}{2}(\rho_{11}-\rho_{22}),\\
\label{eq:3}
\dv{\rho_{11}}{t}&= -i \frac{\varOmega_{\mathrm{d}}}{2} (\rho_{21}-\rho_{12})-\varLambda_{\mathrm{p}} \rho_{11}+R_{31}\rho_{33}+R_{51}\rho_{55},\\
\label{eq:4}
\dv{\rho_{22}}{t}&=i \frac{\varOmega_{\mathrm{d}}}{2} (\rho_{21}-\rho_{12})-\varLambda_{\mathrm{p}} \rho_{22}+R_{42}\rho_{44}+R_{52}\rho_{55},\\
\label{eq:5}
\dv{\rho_{33}}{t}&=\varLambda_{\mathrm{p}} \rho_{11} - (R_{31} + R_{35}) \rho_{33},\\
\label{eq:6}
\dv{\rho_{44}}{t}&=\varLambda_{\mathrm{p}} \rho_{22} - (R_{42} + R_{45}) \rho_{44},\\
\label{eq:7}
\dv{\rho_{55}}{t}&= R_{35} \rho_{33} + R_{45} \rho_{44} - (R_{52} + R_{51}) \rho_{55},
\end{align}
where $\sum_{i}\rho_{ii}=1$.

The steady-state solution yields the total ground-state population $\rho_{\mathrm{g}} = \rho_{11} + \rho_{22}$.
Then we can write the exponential absorption per unit length, $\beta = \sigma D\rho_{\mathrm{g}}$, where $\sigma$ and $D$ are the absorption-cross section and density of NV centres in the diamond crystal, respectively. Note that we define $\beta$ as an exponent of e, while literature values are often given as an exponent of 10.

For a particular $\varLambda_{\mathrm{p}}$ and $\varOmega_{\mathrm{d}}$, the resonant MW driving ($\varDelta_{\mathrm{g}}=0$) evenly distributes the ground-state population of NV centres between $\ket{1}$ and $\ket{2}$ at steady state.
On the other hand, the non-resonant MW driving ($\varDelta_{\mathrm{g}} \ne 0$) unevenly distributes the NV ground-state population, preferring  $\ket{1}$ over $\ket{2}$.
Magnetic sensing based on NV absorption \cite{walsworth2017absorbtion, ahmadi2018nitrogen, dumeige2019infrared} is based on the key fact $R_{45}>R_{35}$.
Thus at a particular $\varLambda_{\mathrm{p}}$ and $\varOmega_{\mathrm{d}}$, more NV centre population reaches the singlet level $\ket{5}$, for resonant MW driving ($\varDelta_{\mathrm{g}}=0$), in comparison with the non-resonant MW driving ($\varDelta_{\mathrm{g}}\ne 0$), where population ends up mainly in state $\ket{1}$ and $\ket{3}$.
Since $\ket{5}$ has a relatively longer lifetime than that of excited-state levels, the ground-state population $\rho_{\mathrm{g}}$ at steady-state for resonant MW driving is lower than that of non-resonant MW driving   \cite{walsworth2017absorbtion, ahmadi2018nitrogen, dumeige2019infrared}.
As a result, the absorption is higher for detuned MW driving ($\varDelta_{\mathrm{g}}\ne 0$) than that of resonant MW driving ($\varDelta_{\mathrm{g}}= 0$).
Since an external magnetic field induces detuning through Zeeman splitting, the absorption is also modified by such an external magnetic field and thus enables magnetic sensing.

\subsection{\label{subsec:2.2}Diamond Raman laser with NV centres as absorbers in the crystal}

We consider a Raman laser generating light from Raman scattering off a diamond crystal containing also NV centres in a Fabry-Perot cavity with pumping at a wavelength below the ZPL (schematic is shown in Figure \ref{fig:1}).
We follow modelling the Raman laser in an approach similar to references \cite{ding2006theoretical, kitzler2015modelling}.
However, the important difference is the consideration of pump absorption, which otherwise is negligible  \cite{kitzler2015modelling}.

We simplify the system with the following assumptions: 
Both mirrors, M1 and M2 are identical, and highly reflective only for the first Raman wavelength, but fully transmitting the pump as well as all higher-order Raman wavelengths, such that lasing occurs only on the first order.
NV centres are uniformly distributed through the diamond volume in the cavity and the cavity is fully filled with the diamond sample. 
We neglect all contributions into the cavity mode from spontaneous Raman emission as well as spontaneous emission from the NV centres, since the stimulated Raman emission dominates above threshold.
We approximate the fundamental Gaussian cavity mode by a cylinder shape, where the base radius of the cylinder is the cavity beam waist.

The Raman gain for the stimulated Raman scattering (SRS) is $g_{\mathrm{r}} I_{\mathrm{p}} I_{\mathrm{r}}$ \cite{penzkofer1979high,pask2003design,ding2006theoretical,kitzler2015modelling},
Where $g_{\mathrm{r}}$ is the plain Raman gain coefficient, $I_{\mathrm{p}}$ is the intensity of the pump beam, and $I_{\mathrm{r}}$ is the intensity of the Raman beam.
Then the depletion of the pump when it propagates through the diamond position $l$ can be written as \cite{kitzler2015modelling},
\begin{equation}
\label{eq:8}
\dv{I_{\mathrm{p}}}{l} = -\frac{\nu_{\mathrm{p}}}{\nu_{\mathrm{r}}}g_{\mathrm{r}} I_{\mathrm{p}}(l) I_{\mathrm{r}} - \beta I_{\mathrm{p}}(l).
\end{equation}
Here, $\beta=\sigma D \rho_g$ represents the pump absorption, $I_{\mathrm{r}}$ is approximated to be uniform throughout the cavity and $I_{\mathrm{p}}(l)$ is a function of length $l$.
Strictly speaking the pump absorption $\beta=\sigma D \rho_g$ is also a function of position $l$, since different pump intensities $I_p(l)$ lead to different levels of saturation and thus ground state population $\rho_g (l)$. Mathematically this dependence manifests itself in the master equation through the pump rate $\varLambda_{\mathrm{p}}(l) = \sigma I_{\mathrm{p}}(l)/h\nu_{\mathrm{p}}$.
This dependence of $\beta$ on length makes equation (\ref{eq:8}), difficult to solve together with a coupled set of  master equations from \eqref{eq:1} to \eqref{eq:7}.
However, based on a relatively weak reduction of the pump in a single pass and in order to separate the temporal and spatial variables in the differential equations, we assume that $\beta$ is a constant over the length of the cavity. 
For this, we make the approximation that $\varLambda_{\mathrm{p}} (l) =  \varLambda_{\mathrm{p}} \sim \sigma I^{0}_{\mathrm{p}}/h\nu_{\mathrm{p}}$, where $I^{0}_{\mathrm{p}}$ is the intensity of the pump entering to the cavity through M1.
Since $I^{0}_{\mathrm{p}}$ is the maximum intensity of the pump wavelength in the cavity, the pump rate that we consider is the maximum value.
The resultant absorption is higher than the actual scenario and as a result the phase noise  to the NV centres' ground state coherence considered is also high.
However, this approximation enables us to make a good estimate of the Raman laser output generation.
The solution of  equation (\ref{eq:8}) gives $I_{\mathrm{p}}(l)$ as,
\begin{equation}
\label{eq:9}
I_{\mathrm{p}}(l) = I^{0}_{\mathrm{p}}\exp{-l\Bigl(\frac{\nu_{\mathrm{p}}}{\nu_{\mathrm{r}}}g_{\mathrm{r}}I_{\mathrm{r}} + \beta \Bigr)}.
\end{equation}
If $l_{\mathrm{c}}$ is the cavity length, then the depleted intensity of the pump when it passes through the diamond sample can be written as,
\begin{equation}
\label{eq:10}
\Delta I_{\mathrm{p}} = I^{0}_{\mathrm{p}}\Bigl[1- \exp{-l_{\mathrm{c}}\Bigl(\frac{\nu_{\mathrm{p}}}{\nu_{\mathrm{r}}}g_{\mathrm{r}}I_{\mathrm{r}} + \beta \Bigr)}\Bigr].
\end{equation}
Following the method in reference \cite{kitzler2015modelling}, the total intensity of stimulated Raman emission generated in a single pass from equation (\ref{eq:10}) can be written as,
\begin{equation}
\label{eq:11}
\Delta I_{\mathrm{r}} = \frac{g_{\mathrm{r}}I_{\mathrm{r}}}{\frac{\nu_{\mathrm{p}}}{\nu_{\mathrm{r}}}g_{\mathrm{r}}I_{\mathrm{r}}+\beta}\Delta I_{\mathrm{p}}\cdot
\end{equation}
The temporal rate of Raman generation is $(c/n)\Delta I_{\mathrm{r}}/l_{\mathrm{c}}$ \cite{ding2006theoretical}, where $c$ and $n$ are the speed of light and the refractive index of the diamond respectively.
Hence, the temporal rate equation for the Raman laser intensity inside the cavity can be written as,
\begin{equation}
\label{eq:12}
\dv{I_{r}}{t}=\frac{c/n}{l_{\mathrm{c}}}\Delta I_{\mathrm{r}}-\kappa_{\mathrm{r}}I_{\mathrm{r}},
\end{equation}
where $t$ represents time and $\kappa_{\mathrm{r}}$ is the cavity loss rate through the mirror transmissions.
Using equations (\ref{eq:10}), (\ref{eq:11}) and (\ref{eq:12}), the $I^{0}_{\mathrm{p}}$ at steady state can be written as,
\begin{equation}
\label{eq:13}
I^{0}_{\mathrm{p}} = \frac{\frac{l_{\mathrm{c}}}{c/n}\kappa_{\mathrm{r}}}{\frac{g_{\mathrm{r}}}{\frac{\nu_{\mathrm{p}}}{\nu_{\mathrm{r}}}g_{\mathrm{r}}I_{\mathrm{r}}+\beta}\Bigl[1- \exp{-l_{\mathrm{c}}(\frac{\nu_{\mathrm{p}}}{\nu_{\mathrm{r}}}g_{\mathrm{r}}I_{\mathrm{r}} + \beta)}\Bigr]}.
\end{equation}
Equation (\ref{eq:13}) gives a relation between the incident pump intensity $I^{0}_{\mathrm{p}}$ and the intra-cavity Raman laser intensity $I_{\mathrm{r}}$. 
The intensity of the Raman laser output ($I^{\mathrm{out}}$) emitted from the cavity can, then be written as,
\begin{equation}
\label{eq:14}
I^{\mathrm{out}}_{\mathrm{r}}= \frac{l_{\mathrm{c}}}{c/n}\kappa_{\mathrm{r}}I_{\mathrm{r}}.
\end{equation}
The pump and the Raman laser output power can be obtained by multiplying respective intensities with the area of the cavity beam ($A_{\mathrm{beam}}$).
Then using equations (\ref{eq:13}) and equation (\ref{eq:14}), we have the pump and laser output relation incorporating NV absorption.
We obtain an analytical solution for $\beta$ using the steady-state solution of the master equations for $\rho_{\mathrm{g}}$ from Section \ref{subsec:2.1}.
This analytical solution is a lengthy one and we do not explicitly write it down here.
Since $\rho_{\mathrm{g}}$ is a function of $\varLambda_{\mathrm{p}}$, $\beta$ is a function of $I^{0}_{\mathrm{p}}$.
Plugging this analytical form for $\beta$ into equation (\ref{eq:13}), we numerically solve the combination of equations (\ref{eq:13}) and (\ref{eq:14}), to obtain the pump power for each Raman laser output power.
Then we numerically study the response of such a laser to external magnetic fields.

\section{\label{sec:level3}Results and discussion}

We consider a micro-cavity Raman laser that leads to strong light intensity and Raman scattering thereby reducing the laser threshold.
An interesting candidate for this would be fibre cavities \cite{hunger2010fiber, nair2020amplification}.
We consider the length of the diamond sample $l_{\mathrm{c}}$ as \SI{100}{\micro\meter} and the beam waist radius as \SI{5}{\micro\meter}.
Since we are considering a microcavity, for strong pump absorption, we consider a high density of NV centres, about 10 ppm ($1.77 \times 10^{18}$\,cm$^{-3}$). 
Diamonds with even higher density of NV centres have been shown in experiment \cite{acosta2009diamonds, subedi2019laser, Capelli2019}.
The absorption cross-section of the NV centre for the considered pump wavelength ($\sigma$) is approximately 1.3 $\times 10^{-17}$cm$^{2}$, which we obtain from scaling the literature value for 532~nm with a measured absorption spectrum to the wavelength in question \cite{Kern2017}.
The Raman gain ($g_{\mathrm{r}}$) for the choice of our wavelength is around 14.75~cm/GW \cite{sabella2015pump, mildren2013optical}.

\subsection{\label{subsec:3.1}Raman laser LTM}
\begin{figure}
\centering
\includegraphics[scale=1]{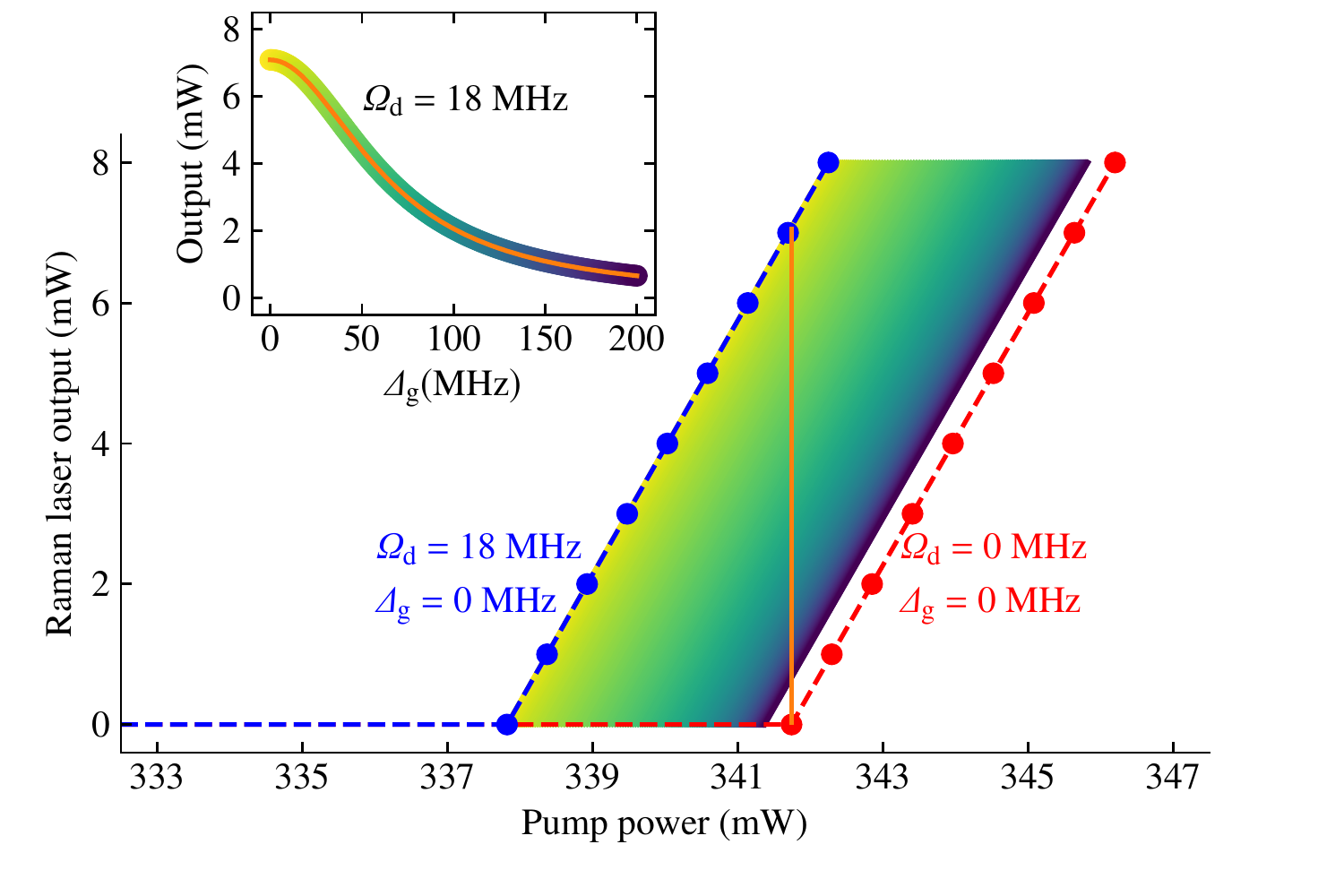}
\caption{Raman laser output power in milli-Watt (mW) as a function of pump power for different MW driving conditions and for a cavity loss rate of $\kappa_{\mathrm{r}}$ = 75~MHz.
The blue dots are the numerically obtained solutions of Raman-laser output power for resonant MW driving ($\varDelta_{\mathrm{g}} = 0$~MHz) with $\varOmega_{\mathrm{d}}=18$~MHz.
The red dots are the solutions when no MW field driving is present ($\varOmega_{\mathrm{d}}=0$~MHz)
The dashed curves are interpretation laser curves considering the Raman laser output power is zero when the pump power is zero.
The gradient shaded region is formed by joining the line plots of numerical solutions corresponding to different $\varDelta_{\mathrm{g}}$ from 0 to 200~MHz for $\varOmega_{\mathrm{d}}=18$~MHz.
For a pump power equal to the threshold pump power of the MW-off laser curve, the Raman-laser output changes through the orange line cut.
Raman-laser output as a function of detuning is shown in the inset. 
The data points are shown in the corresponding gradient colour in the main figure.
The orange curve in the inset is the  interpolation function plot.
}
\label{fig:2}
\end{figure}

An example of numerical solutions of the Raman laser output as a function of pump laser power, for different MW driving conditions, is shown in Figure \ref{fig:2}.
Since the absorption by NV centres is less for the resonant MW driving ($\varDelta_{\mathrm{g}}$ = 0) than for detuned MW driving($\varDelta_{\mathrm{g}} \ne 0$) or MW-off case ($\varOmega_{\mathrm{d}}$ = 0), the resonant case reaches the threshold for lower pump powers than the detuned case or MW-off case. 
As the detuning $\varDelta_{\mathrm{g}}$ increases, the laser curves tends towards that of the MW-off case.
The threshold corresponding to the MW-off case is fundamentally limited by the number of NV centres in the cavity volume and their internal transition rates.

\begin{figure}
\centering
\subfigure{\label{fig:3.1}\includegraphics[scale=0.5]{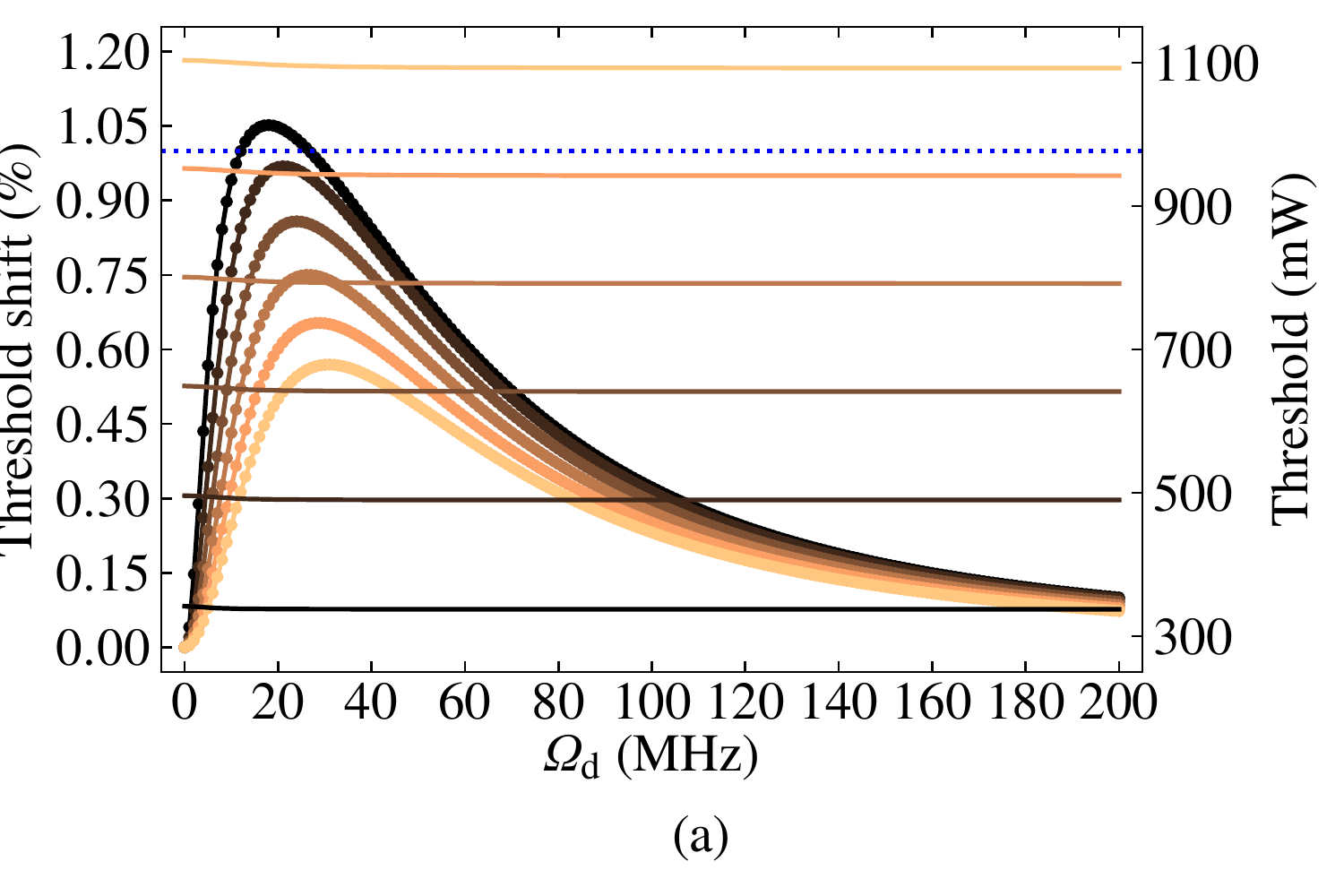}}
\subfigure{\label{fig:3.2}\includegraphics[scale=0.5]{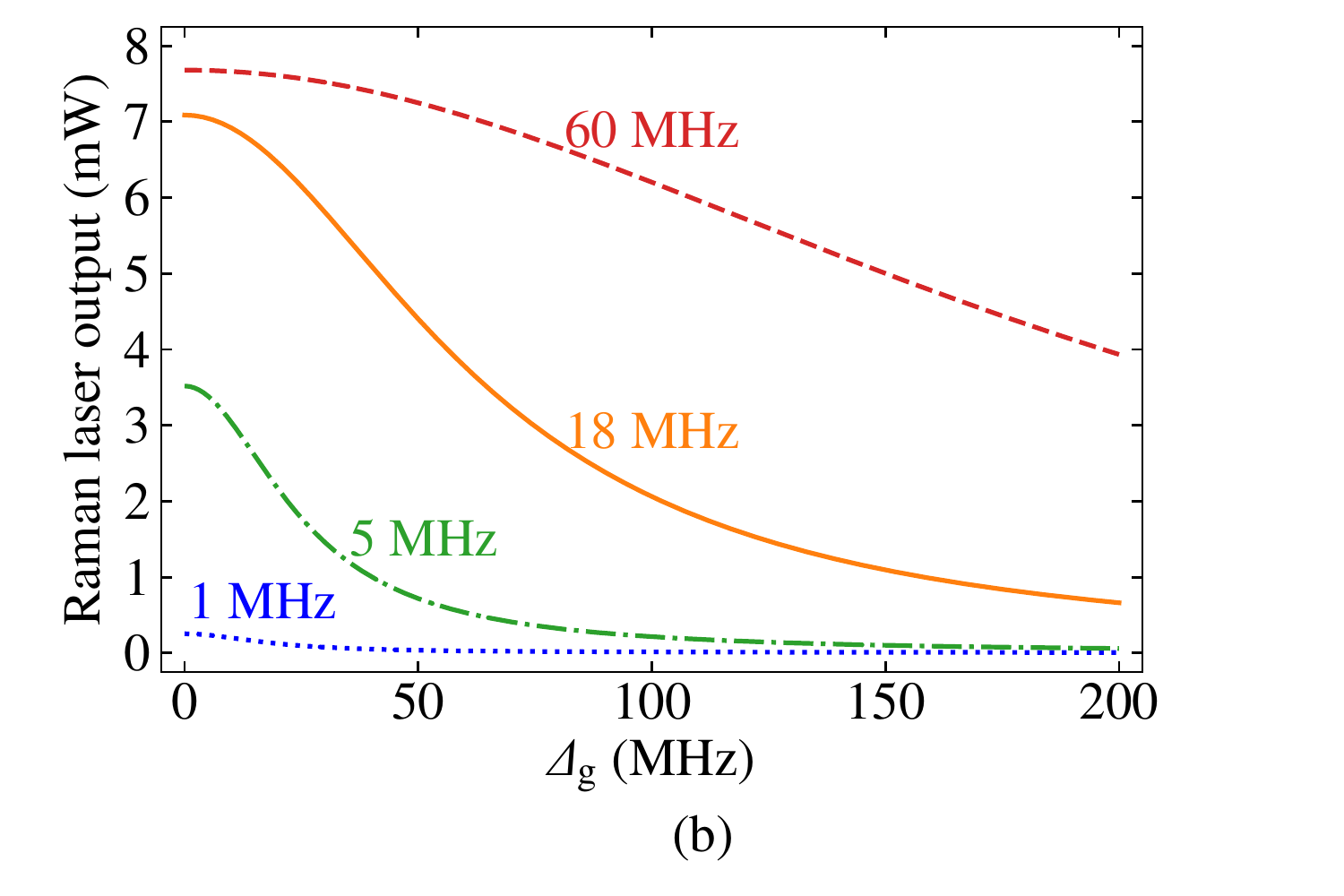}}
\subfigure{\label{fig:3.3}\includegraphics[scale=0.5]{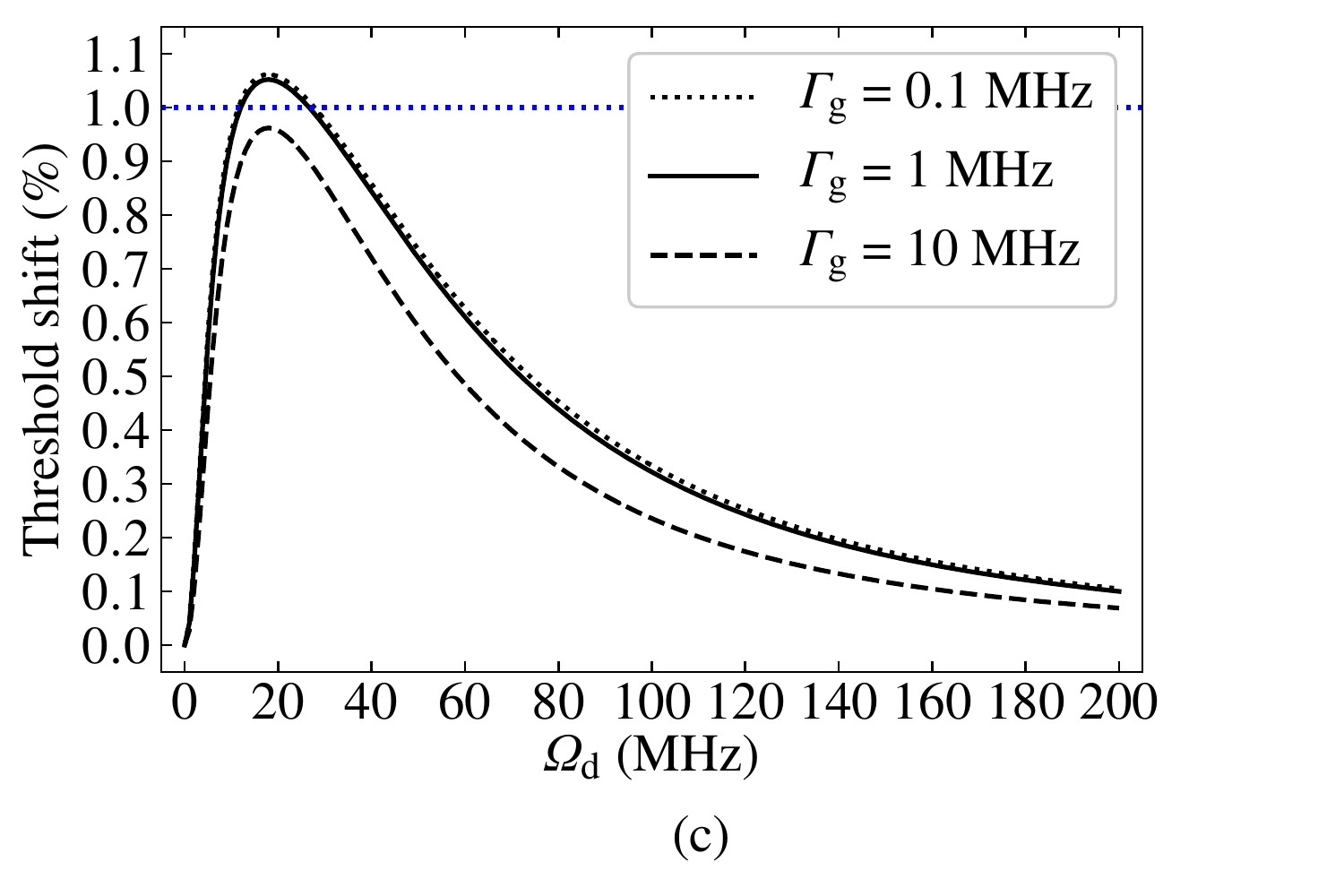}}
\subfigure{\label{fig:3.4}\includegraphics[scale=0.5]{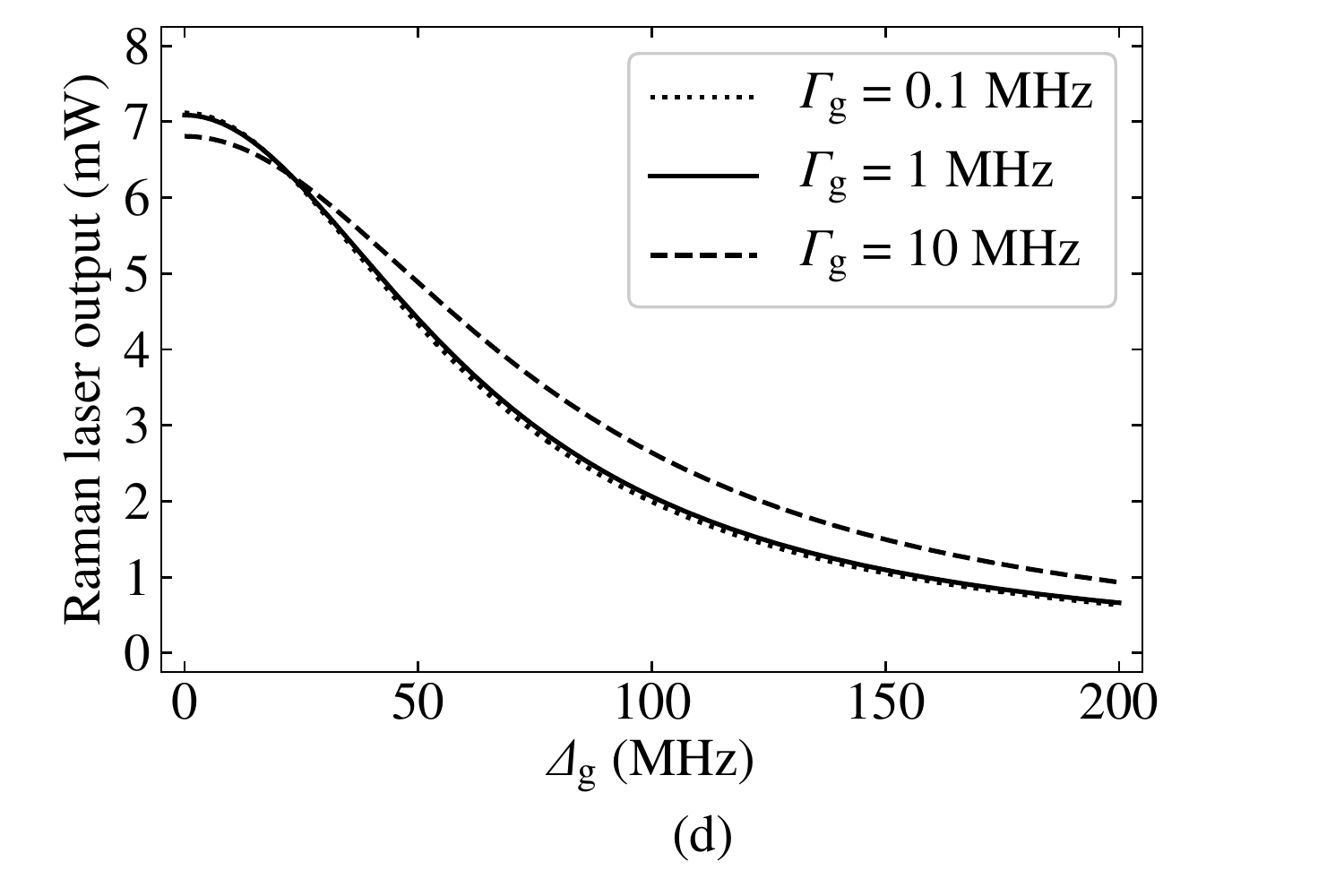}}
\caption{
(a) The shift in threshold pump power for detuned laser curves ($\varDelta_{\mathrm{g}}$ = 200~MHz) with respect to that of resonant laser curves ($\varDelta_{\mathrm{g}}$ = 0~MHz) as a percentage of threshold pump power of resonant laser curves  (numerical solutions shown with dots).
The dotted blue line corresponds to 1~$\%$ shift in the threshold.
In addition, we show the threshold pump power of resonant laser curves (line plots of the numerical solutions) as a function of $\varOmega_{\mathrm{d}}$ for different cavity loss-rates $\kappa_{\mathrm{r}}$.
The plots are for $\kappa_{\mathrm{r}}$ ranging from 75~MHz (black colour plots) to 250~MHz in 35~MHz steps.
$\varGamma_{\mathrm{g}}$ is 1~MHz.
(b) Raman-laser output power as a function of $\varDelta_{\mathrm{g}}$ for $\kappa_{\mathrm{r}}$ = 75~MHz, $\varGamma_{\mathrm{g}}$ = 1~MHz, and for different values of $\varOmega_{\mathrm{d}}$.
(c) Threshold shift as in (a), now as a function of $\varOmega_{\mathrm{d}}$ for $\kappa_{\mathrm{r}}$ = 75~MHz, and for different values of $\varGamma_{\mathrm{g}}$.
(d) Raman laser output as a function of $\varDelta_{\mathrm{g}}$ for $\kappa_{\mathrm{r}}$ = 75~MHz, $\varOmega_{\mathrm{d}}$ = 18~MHz, and for different values of $\varGamma_{\mathrm{g}}$. The values of $\varGamma_{\mathrm{g}}$ corresponding to each curve is denoted in the respective colour.
}
\label{fig:3}
\end{figure}

We can apply the LTM method to detect the external magnetic field by assuming that the MW frequency for driving the NV centre is constant at a fixed value and the detuning is induced by some external magnetic field \cite{jeske2016laser}.
For this, we set the pump power to $P'_{\mathrm{p}}$, which is the upper threshold value (red curve) in the main Figure \ref{fig:2} \cite{dumeige2019infrared}.
Then looking at the change in the Raman laser output for resonant MW driving corresponding to $P'_{\mathrm{p}}$ pump power, we can infer the detuning and thereby the magnetic field present, since $B_{\mathrm{DC}} = \varDelta_{\mathrm{g}}/\gamma_{e}$, where $1/\gamma_{e} = 5.68\times10^{-12}$~T/Hz \cite{jeske2016laser}.
For the rest of the study we consider $\varDelta_{\mathrm{g}}$ between 0~MHz and 200~MHz for numerical estimations.
The output power of the Raman laser as a function of the detuning then forms a peak around zero, as shown in the inset figure of Figure \ref{fig:2}, since at resonance the absorption is the lowest.
For $\varDelta_{\mathrm{g}} < 0$, the curve in the inset mirrors with respect to $\varDelta_{\mathrm{g}} = 0$.
Since we require only $|B_{\mathrm{DC}}|$ for estimating the DC magnetic field, we focus on the positive values of $\varDelta_{\mathrm{g}}$.
The peak width of this curve is set by the power broadening and phase noise that arises due to the Rabi driving $\varOmega_{\mathrm{d}}$, dephasing rate $\varGamma_{\mathrm{g}}$, and laser pumping $\varLambda_{\mathrm{p}}$, to the NV ground state coherence.
The laser pumping $\varLambda_{\mathrm{p}}$ is determined by the threshold pump powers, which basically depends on the cavity loss rate $\kappa_{\mathrm{r}}$, for fixed number of NV centres and cavity volume as we considered.
We explore the effects of $\varOmega_{\mathrm{d}}$, $\varGamma_{\mathrm{g}}$, and $\kappa_{\mathrm{r}}$ in Figure \ref{fig:3}.

Since we are fixing the pump power to $P'_{\mathrm{p}}$ and looking at the Raman laser output change, any undesired fluctuations in the pump power is expected to produce noise in the Raman laser output power masking the magnetic-field signal we want to detect. 
Thus in experiments, it is important to work with a pump laser that is stabilized for power and/or monitor its fluctuations, e.g. with a balanced photodiode detection.
As a result it is interesting to maximise the threshold shift between resonant ($\varDelta_{\mathrm{g}}$ = 0~MHz) and off-resonant driving ($\varDelta_{\mathrm{g}}$ = 200~MHz).
We therefore explore the threshold behaviour of the Raman laser as a function of Rabi driving $\varOmega_{\mathrm{d}}$ and cavity loss rate $\kappa_{\mathrm{r}}$ in Figure~\ref{fig:3}(a).
The threshold pump power for resonant driving ($\varDelta_{\mathrm{g}}$ = 0~MHz) is almost constant between $\varOmega_{\mathrm{d}}$ = 0~MHz and $\varOmega_{\mathrm{d}}$ = 200~MHz, but almost linearly increases when increasing the cavity loss rate (decreasing the finesse of the cavity).
On the other hand, the threshold shift between resonant ($\varDelta_{\mathrm{g}}$ = 0~MHz) and off-resonant driving ($\varDelta_{\mathrm{g}}$ = 200~MHz) as a percentage of the resonant threshold pump power shows a characteristic maximum (i.e.~optimum) between $\varOmega_{\mathrm{d}}$ = 0~MHz and $\varOmega_{\mathrm{d}}$= 200~MHz.
The finesse of the cavity corresponding to  $\kappa_{\mathrm{r}}$ = 75~MHz is around 52360 and we get more than 1\% threshold shift for an optimum $\varOmega_{\mathrm{d}}$ value of 18 MHz.
The $P'_{\mathrm{p}}$ for this case is around 341.74~mW, which corresponds to a $\varLambda_{\mathrm{p}}$ around 17.64~MHz.
For a cavity loaded with a pure diamond sample of around \SI{10}{\micro\meter} thickness, a cavity finesse of around 17000 has been demonstrated before \cite{janitz2015fabry}.
Due to the high finesse of approximately 52000, the intra-cavity power for around 5~mW output (multiplying with finesse/$\pi$) can be roughly estimated to be around 83~W.
Thus one has to be careful about the damage threshold of the mirror coatings of the cavity mirrors. 

In order to understand the optimised effect in the threshold shift, we consider the $\kappa_{\mathrm{r}}$ = 75~MHz curve in Figure \ref{fig:3}(a) and plot the Raman laser output curve as a function of $\varDelta_{\mathrm{g}}$ for different values of $\varOmega_{\mathrm{d}}$ as shown in Figure \ref{fig:3}(b).
Figure \ref{fig:3}(b), shows that increasing  $\varOmega_{\mathrm{d}}$ improves the Raman laser output, but broadens the resonance peak.
The resonance peak broadening for higher $\varOmega_{\mathrm{d}}$ is expected due to the significant contribution into phase noise to the ground state coherence by $\varOmega_{\mathrm{d}}$.
The effect of $\varGamma_{\mathrm{g}}$ explored in Figure \ref{fig:3}(c) and \ref{fig:3}(d), shows that improving $\varGamma_{\mathrm{g}}$ by an order of magnitude from 1~MHz does not really improve anything. 
On the other hand, increasing $\varGamma_{\mathrm{g}}$ by an order of magnitude changes both threshold shift and the Raman laser output resonance peak width.
We attribute this effect to the fact that $\varLambda_{\mathrm{p}}$ around 17.64~MHz is the dominant source of phase noise to the ground state coherence and increasing the $\varGamma_{\mathrm{g}}$ by an order of magnitude from 1~MHz makes $\varGamma_{\mathrm{g}}$ comparable to $\varLambda_{\mathrm{p}}$.

\subsection{\label{subsec:3.2}Sensitivity of the present magnetometry scheme and its optimisation}

\begin{figure}
\centering
\subfigure
{\label{fig:1.1}\includegraphics[scale=0.5]{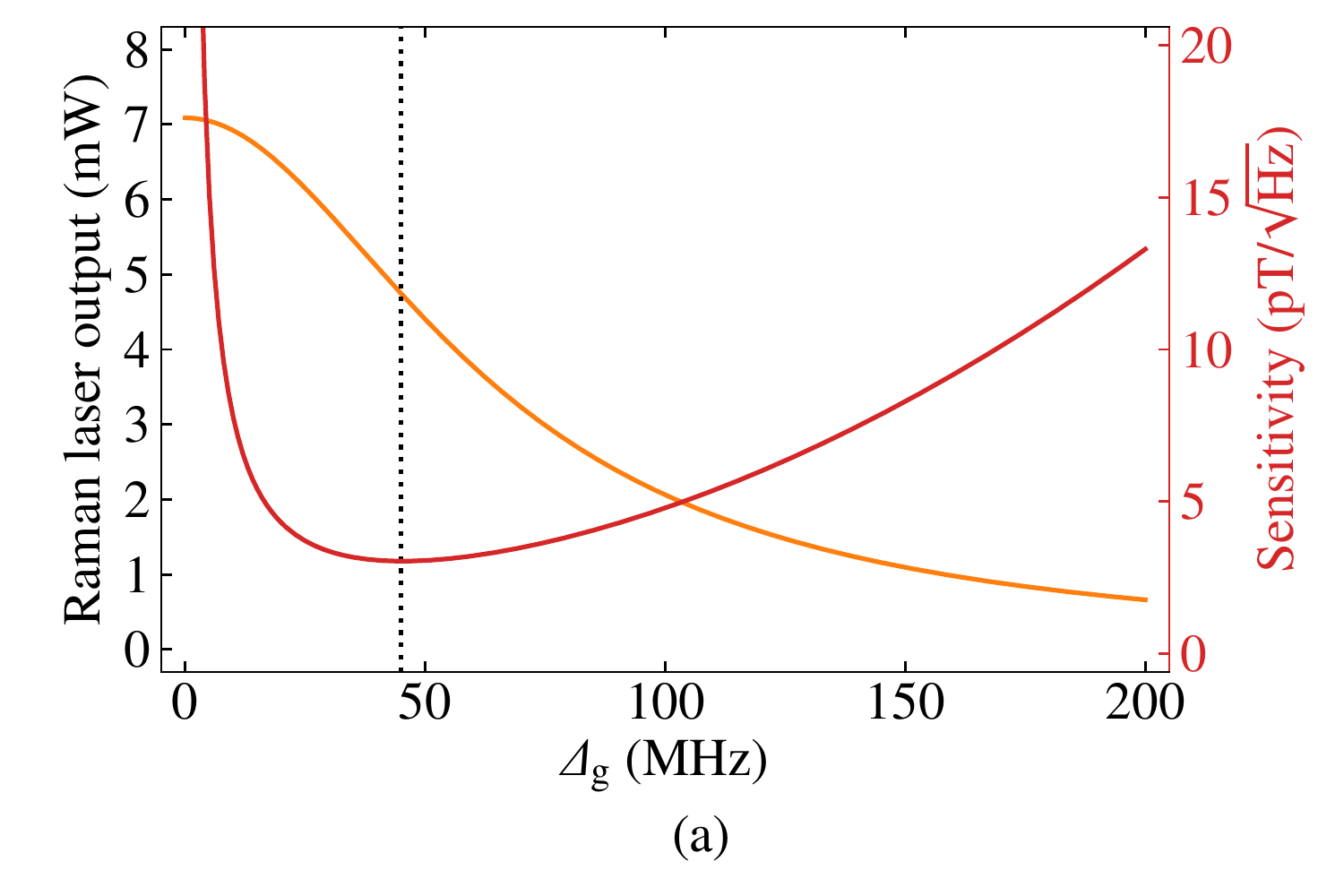}
}
\subfigure
{\label{fig:2.2}\includegraphics[scale=0.5]{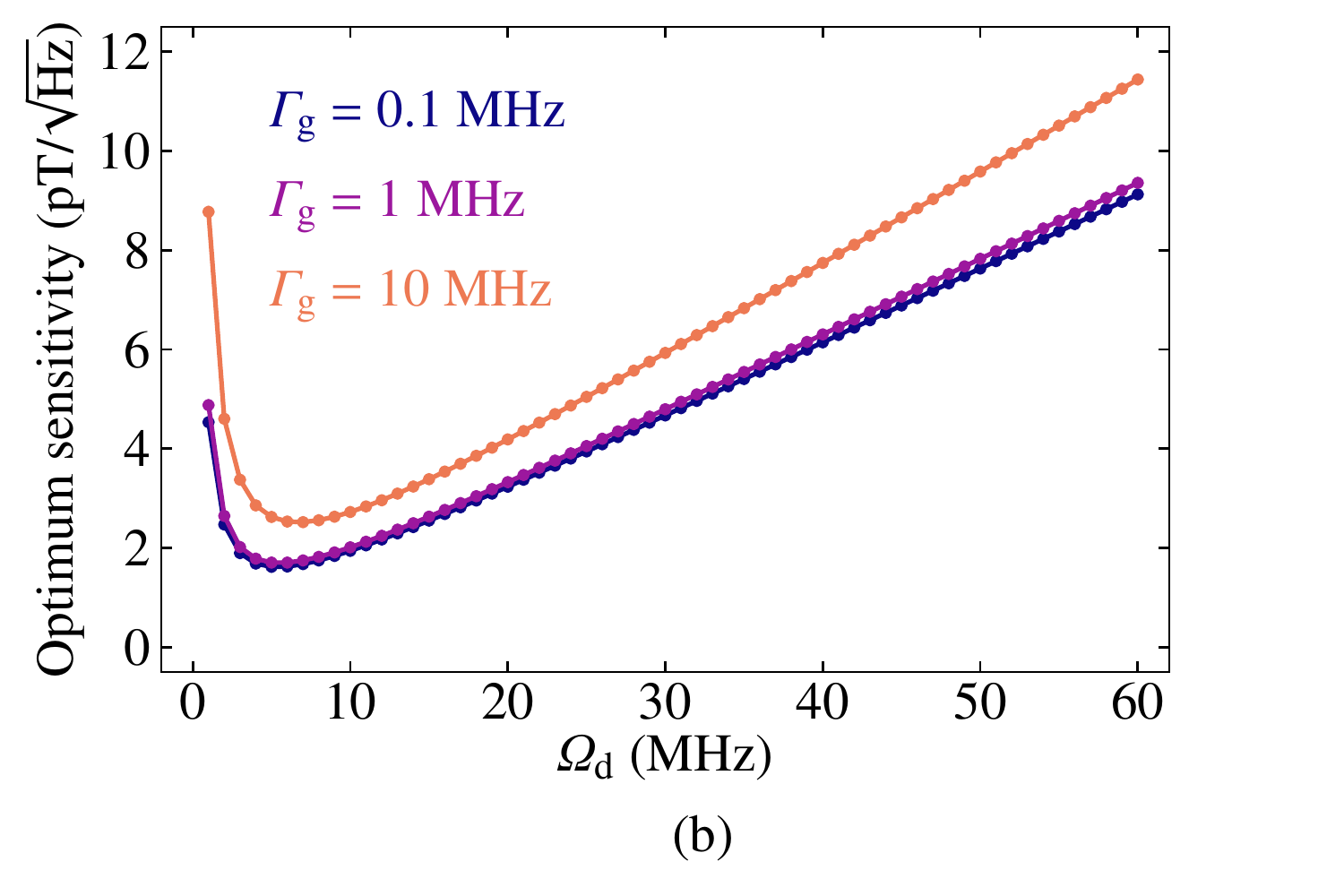}}
\caption{(a) The orange solid curve was obtained by an interpolation fit to the Raman laser output as a function of detuning as shown in Figure \ref{fig:2}.
The solid red curve is the resulting sensitivity as a function of the detuning, which is related to the slope of the orange curve.
As the detuning tends to zero, the sensitivity increases as the slope in the orange curve tends to zero.
The y-axis for the sensitivity is shown in red colour.
The black dotted vertical line is at the detuning corresponding to the minimum sensitivity.
(b) The minimum sensitivity as a function $\varOmega_{\mathrm{d}}$ for different dephasing rates $\varGamma_{\mathrm{g}}$, as indicated in the figure legends in corresponding colours of the plots.}
\label{fig:4}
\end{figure}

The dominant and limiting noise in the magnetic field sensing using the present scheme is expected to be the photon-shot noise and thus we estimate a photon-shot noise limited magnetic field sensitivity of the system \cite{degen2017quantum, Rondin2014, jeske2016laser}.
If the Raman laser output ($P_{\mathrm{r}}^{\mathrm{out}}$) varies by an infinitesimally amount $\mathrm{d}P_{\mathrm{r}}^{\mathrm{out}}$ corresponding to an infinitesimally small change $\mathrm{d}B_{\mathrm{DC}}$ in the DC magnetic field, for a finite measurement time $\mathrm{d}t$, then the photon-shot noise and the signal photon number that informs about the magnetic field can be written as  $\sqrt{(P_{\mathrm{r}}^{\mathrm{out}}/h\nu_{\mathrm{r}})\mathrm{d}t}$ and $(\mathrm{d}P_{\mathrm{r}}^{\mathrm{out}}/h\nu_{\mathrm{r}})\mathrm{d}t$ respectively.
Then the photon-shot noise limited sensitivity ($\eta_{\mathrm{DC}}=|\mathrm{d}B_{\mathrm{DC}}|\sqrt{\mathrm{d}t}$) to the DC magnetic field $B_{\mathrm{DC}}$ can be written from equating the signal photon number to the photon shot noise as \cite{degen2017quantum, Rondin2014}, $\eta_{\mathrm{DC}}~=~\sqrt{h\nu_{\mathrm{r}}P_{\mathrm{r}}^{\mathrm{out}}}~\times~|\dv*{B_{\mathrm{DC}}}{P_{\mathrm{r}}^{\mathrm{out}}|}$.
The derivative $|\dv*{B_{\mathrm{DC}}}{P_{\mathrm{r}}^{\mathrm{out}}|}$ can be obtained as the inverse of the slope of the laser output curve as a function of detuning using the gyromagnetic ratio ($\gamma_{e}$) of the NV centre.
We estimate and optimise the minimum sensitivity of the present magnetometry scheme as shown in Figure \ref{fig:4}.

The sensitivity corresponding to the Raman laser presented in Figure \ref{fig:2} as a function of detuning is shown in Figure \ref{fig:4}(a).
To better determine the slope of the Raman laser output curve as a function of detuning, we fit the numerical solutions with the interpolation function (orange colour curve).
The derivative of the obtained interpolation function is used to estimate the sensitivity.
Close to zero detuning the laser is not sensitive to the external magnetic field since the slope of the laser output goes to zero.
Thus the optimal magnetic field sensitivity is at a specific detuning value simply found by minimising the expression $\eta_{\mathrm{DC}}=\sqrt{h\nu_{\mathrm{r}}P_{\mathrm{r}}^{\mathrm{out}}}~\times~|\dv*{B_{\mathrm{DC}}}{P_{\mathrm{r}}^{\mathrm{out}}|}$, i.e.~maximising the slope of the orange curve in Fig.~\ref{fig:4}(a) divided by the square root of the power output $(|\dv*{P_{\mathrm{r}}^{\mathrm{out}}}{\varDelta_{\mathrm{g}}|})/\sqrt{P_{\mathrm{r}}^{\mathrm{out}}}$.

We show the minimum sensitivity  as a function of Rabi frequency $\varOmega_{\mathrm{d}}$ for different dephasing rates $\varGamma_{\mathrm{g}}$ in Figure \ref{fig:4}~(b). 
Reducing the $\varOmega_{\mathrm{d}}$, we can improve the sensitivity down to a few pT/$\sqrt{\mathrm{Hz}}$, but the threshold shift is also reduced. 
While a smaller $\varGamma_{\mathrm{g}}$ value close to 1~MHz provides better sensitivity, anything below this value does not improve the sensitivity much.
The minimum sensitivity for $\varGamma_{\mathrm{g}}$ = 0.1~MHz is around \SI{1.62}{\pico\tesla/\sqrt{\hertz}} at $\varOmega_{\mathrm{d}}$ around 5~MHz.
Below $\varOmega_{\mathrm{d}}$ around 5~MHz the sensitivity increases, since the contrast of the Raman laser output between the resonant and non-resonant output reduces significantly and thereby the slope is reduced (see Figure \ref{fig:3}(b)).

To estimate the sensitivity advantage of a diamond Raman LTM cavity we benchmark our scheme with conventional ODMR NV magnetometry.
In conventional ODMR, the volume defined sensitivity ($\eta_{\mathrm{DC}} \times \sqrt{\text{sensing volume}}$), with a diamond sample having 2 ppm NV density and a comparable dephasing rate is about 100-$\SI{200}{\nano\tesla\sqrt{{\micro\meter}^{3}/\hertz}}$ (see figure 10.F \cite{levine2019principles}).
The number of NV centres in this conventional magnetomerty sensing volume would be equal to the number of NV centres that we considered in our cavity, if its sensing volume is 5 times larger than our cavity volume ($7.9 \times 10^{-6}\SI{}{{\milli\meter}^3}$).
Then the sensitivity, $\eta_{\mathrm{DC}}$ of the conventional magnetometry is roughly $\SI{1}{\nano\tesla/\sqrt{\hertz}}$, which is three orders of magnitude worse than the sensitivities we estimated in the present work. Thus we can conclude that comparing the two techniques with both CW laser and CW microwaves, the absorptive diamond Raman laser cavity in our scheme results in a sensitivity advantage of three orders of magnitude.

The sensitivity of conventional NV magnetometry can further be improved by switching from CW techniques to pulsed and/or lock-in techniques.
Focusing only on the sensitivity value, an experimental DC sensitivity with NV centres of about 15~pT/$\sqrt{\mathrm{Hz}}$ \cite{barry2016optical} was demonstrated in 2016. Very recently using a magnetic flux concentrator this was improved to about 0.9~pT/$\sqrt{\mathrm{Hz}}$ \cite{fescenko2020diamond} and using improved pulsed sequences \cite{zhang2020diamond} this has been further reduced to about 0.5~pT/$\sqrt{\mathrm{Hz}}$. In principle these advantages such as a flux concentrator could be combined with LTM schemes. However, in order to combine pulsed schemes with a laser read-out of NV centres, questions of cavity build-up times and their interference with the read-out times arise. Dynamical calculations are therefore needed to discuss sensitivities of such combinations and the steady-state solutions presented in this paper are not sufficient. 

Comparing our scheme to other LTM schemes, naively we can say that the minimum sensitivity we obtain, is comparable to the 0.7~pT/$\sqrt{\mathrm{Hz}}$, projected DC sensitivity for IR laser threshold magnetometry in reference \cite{dumeige2019infrared}, and is a few orders of magnitude worse than the projected one for the NV laser threshold magnetometry \cite{jeske2016laser}.
The dephasing rates considered in these works fall within the values we consider in this work.
However, a proper comparison should consider the total number of NV centres involved, i.e., the diamond volume and the density of NV centres in different schemes. For application advantages, such as low-power operation and fibre-coupling we chose a microcavity with a diamond sensing volume of $\pi\times(\SI{5}{\micro\meter})^{2} \times \SI{100}{\micro\meter}=7.9 \times 10^{-6}\SI{}{{\milli\meter}^3}$. The IR laser threshold magnetometry scheme assumed a diamond volume that is roughly two orders of magnitude larger ($\pi\times(\SI{50}{\micro\meter})^{2}\times\SI{100}{\micro\meter}$) and the NV density larger by a factor of 2.8 \cite{dumeige2019infrared}. 
The active NV lasing scheme assumed 1mm$^3$ of diamond sensing volume with an NV density larger by a factor 1.6 than the present case.

In LTM schemes, the optical cavity parameters can influence the laser output and threshold shift and thus sensitivity. These are additional parameters compared to conventional NV magnetometry, where the linewidth is typically defined by the NV diamond material and the only corresponding setup parameter is the photon collection efficiency. These additional parameters can on the one hand be used to adapt a sensor to serve particular requirements such as wide dynamic range versus sensitivity. On the other hand they can make performance fabrication-dependent and thus require more calibration. However, we point out here, that the cavity and fabrication has no influence on the fundamental NV centre's magnetic resonance frequency or its shift with the external magnetic field, purely defined by the fundamental gyromagnetic ratio. Thus the main fundamental advantage of NV centres over other magnetometer technologies, often labelled 'calibration-free' measurements, is preserved.

To reach this sensitivity, a number of underlying assumptions, as discussed earlier, have to be met.
The main assumptions are that frequency and power are stable for the pump wavelength, and the NV centres are charge stable.
The charge state switching of the NV centre by the pump wavelength can have detrimental effects on the estimated sensitivities. 
There is a growing interest in controlling the charge state of the NV centres \cite{alkahtani2020charge}.
If due to the chosen wavelength of \SI{620}{\nano\meter} partial photoionization or charge-state switching of the NV centres \cite{aslam2013photo, manson2018nv} occurs, this can be mitigated by adding an additional green laser for re-pumping, the neutral charge state to the negative charge state \cite{aslam2013photo}.
Intrinsically, the charge-state switching is indirectly spin dependent due to the spin-dependent shelving into the singlet state.
On the otherhand, the experimental study in reference \cite{roberts2019spin} reports that the charge-state switching of the NV centres in nanodiamonds is directly spin-dependent.
However, an accurate estimation of the response of the Raman laser against magnetic field and its sensitivity, would require inclusion of the effect of charge-state switching at the wavelength in the present study.
However, this is well beyond the scope of the present study and demands more experimental as well as theoretical work in the future.

\section{\label{sec:level4}Conclusions}

We have shown that a low threshold single pass pump diamond Raman laser with absorption of only the pump wavelength by MW driven NV centres in the same diamond crystal can be used to detect external magnetic fields.
The choice of a micro-cavity with high finesse for the Raman wavelength reduces the threshold powers well below NV saturation powers and thus could investigate the role of NV centres, though one has to be careful about the damage thresholds.
We have explored the response of such a Raman laser as a function of incoherent laser pumping, MW driving strength, and intrinsic dephasing rate. We find that the laser threshold shifts between on- and off-resonant MW driving of the NV centres, which allows for sensing. This threshold shift is maximised by an improved optical cavity, i.e.~lower cavity loss rate and further maximised for a specific optimal Rabi-driving, which is around $\varOmega_{\mathrm{d}}=18$~MHz for our parameters.
Furthermore, we have estimated the DC magnetic field sensitivity and optimised it, where reducing the intrinsic NV centre's dephasing rate improves the sensitivity, essentially until the dephasing rate approaches the NV centres' incoherent laser pump rate.
We obtain a minimum sensitivity of down to a few pT/$\sqrt{\mathrm{Hz}}$ and find that for comparable sensing volume, NV density and coherence time our scheme gives an advantage of three orders of magnitude in sensitivity compared with conventional ODMR.
The sensitivity range projected in this study is comparable with the projected DC magnetic sensitivity of other LTM magnetometry with NV centres reported  \cite{dumeige2019infrared, jeske2016laser}, but the conceptual simplicity of our scheme can attract potential applications, especially since it takes advantage of the well-established, visible wavelength Raman lasing properties also provided by diamond.

\section*{Acknowledgments}
Sarath Raman Nair acknowledges that this work was done while holding the International Macquarie University Research Excellence Scholarship (iMQRES) (Allocation number 2014108) and also acknowledges the EQUS Centre Collaboration Award. Lachlan J. Rogers is the recipient of an Australian Research Council Discovery Early Career Award (project number DE170101371). This research was supported by the Australian Research Council Centre of Excellence for Engineered Quantum Systems (EQUS, CE170100009). Jan Jeske acknowledges funding from the German federal ministry for education and research Bundesministerium f\"ur Bildung und Forschung (BMBF) under grant number  13XP5063. Andrew D. Greentree acknowledges the support of an ARC Future Fellowship (Grant No. FT160100357)

\section*{Reference}

\bibliographystyle{iopart-num}
\bibliography{ref}

\end{document}